\documentclass[fleqn,a4paper,12pt,twoside]{article}

\usepackage{espcrc1}
\usepackage{epsfig}
\newcommand{\GeV}{\mbox{$A$GeV}}
\newcommand{\fig}[1]{Fig.~\ref{#1}}
\newcommand{\dEdx}{\ensuremath{dE/dx}}
\newcommand{\omeg}{\ensuremath{\Omega^-}}
\newcommand{\aomeg}{\ensuremath{\bar{\Omega}^+}}
\title{Recent results on spectra and yields from NA49}

\author{M. van Leeuwen for the NA49 collaboration\footnote{Presented
    at Quark Matter 2002, Nantes, France.} \\
\vspace{0.3cm}
{\footnotesize
S.V.~Afanasiev$^{9}$,T.~Anticic$^{21}$, B.~Baatar$^{9}$,D.~Barna$^{5}$,
J.~Bartke$^{7}$, R.A.~Barton$^{3}$, M.~Behler$^{15}$,
L.~Betev$^{10}$, H.~Bia{\l}\-kowska$^{19}$, A.~Billmeier$^{10}$,
C.~Blume$^{8}$, C.O.~Blyth$^{3}$, B.~Boimska$^{19}$, M.~Botje$^{1}$,
J.~Bracinik$^{4}$, R.~Bramm$^{10}$, R.~Brun$^{11}$,
P.~Bun\v{c}i\'{c}$^{10,11}$, V.~Cerny$^{4}$, O.~Chvala$^{17}$,
J.G.~Cramer$^{18}$, P.~Csat\'{o}$^{5}$, P.~Dinkelaker$^{10}$,
V.~Eckardt$^{16}$, P.~Filip$^{16}$,
H.G.~Fischer$^{11}$, Z.~Fodor$^{5}$, P.~Foka$^{8}$, P.~Freund$^{16}$,
V.~Friese$^{8,15}$, J.~G\'{a}l$^{5}$,
M.~Ga\'zdzicki$^{10}$, G.~Georgopoulos$^{2}$, E.~G{\l}adysz$^{7}$, 
S.~Hegyi$^{5}$, C.~H\"{o}hne$^{15}$, G.~Igo$^{14}$,
P.G.~Jones$^{3}$, K.~Kadija$^{11,21}$, A.~Karev$^{16}$,
V.I.~Kolesnikov$^{9}$, T.~Kollegger$^{10}$, M.~Kowalski$^{7}$, 
I.~Kraus$^{8}$, M.~Kreps$^{4}$, M.~van~Leeuwen$^{1}$, 
P.~L\'{e}vai$^{5}$, A.I.~Malakhov$^{9}$, S.~Margetis$^{13}$,
C.~Markert$^{8}$, B.W.~Mayes$^{12}$, G.L.~Melkumov$^{9}$,
C.~Meurer$^{10}$,
A.~Mischke$^{8}$, M.~Mitrovski$^{10}$, 
J.~Moln\'{a}r$^{5}$, J.M.~Nelson$^{3}$,
G.~P\'{a}lla$^{5}$, A.D.~Panagiotou$^{2}$,
K.~Perl$^{20}$, A.~Petridis$^{2}$, M.~Pikna$^{4}$, L.~Pinsky$^{12}$,
F.~P\"{u}hlhofer$^{15}$,
J.G.~Reid$^{18}$, R.~Renfordt$^{10}$, W.~Retyk$^{20}$,
C.~Roland$^{6}$, G.~Roland$^{6}$, A.~Rybicki$^{7}$, T.~Sammer$^{16}$,
A.~Sandoval$^{8}$, H.~Sann$^{8}$, N.~Schmitz$^{16}$, P.~Seyboth$^{16}$,
F.~Sikl\'{e}r$^{5}$, B.~Sitar$^{4}$, E.~Skrzypczak$^{20}$,
G.T.A.~Squier$^{3}$, R.~Stock$^{10}$, H.~Str\"{o}bele$^{10}$, T.~Susa$^{21}$,
I.~Szentp\'{e}tery$^{5}$, J.~Sziklai$^{5}$,
T.A.~Trainor$^{18}$, D.~Varga$^{5}$, M.~Vassiliou$^{2}$,
G.I.~Veres$^{5}$, G.~Vesztergombi$^{5}$,
D.~Vrani\'{c}$^{8}$, S.~Wenig$^{11}$, A.~Wetzler$^{10}$, C.~Whitten$^{14}$,
I.K.~Yoo$^{8,15}$, J.~Zaranek$^{10}$, J.~Zim\'{a}nyi$^{5}$\\
\vspace{0.3cm}
\noindent
$^{1}$NIKHEF, Amsterdam, Netherlands. \\
$^{2}$Department of Physics, University of Athens, Athens, Greece.\\
$^{3}$Birmingham University, Birmingham, England.\\
$^{4}$Comenius University, Bratislava, Slovakia.\\
$^{5}$KFKI Research Institute for Particle and Nuclear Physics, Budapest, Hungary.\\
$^{6}$MIT, Cambridge, USA.\\
$^{7}$Institute of Nuclear Physics, Cracow, Poland.\\
$^{8}$Gesellschaft f\"{u}r Schwerionenforschung (GSI), Darmstadt, Germany.\\
$^{9}$Joint Institute for Nuclear Research, Dubna, Russia.\\
$^{10}$Fachbereich Physik der Universit\"{a}t, Frankfurt, Germany.\\
$^{11}$CERN, Geneva, Switzerland.\\
$^{12}$University of Houston, Houston, TX, USA.\\
$^{13}$Kent State University, Kent, OH, USA.\\
$^{14}$University of California at Los Angeles, Los Angeles, USA.\\
$^{15}$Fachbereich Physik der Universit\"{a}t, Marburg, Germany.\\
$^{16}$Max-Planck-Institut f\"{u}r Physik, Munich, Germany.\\
$^{17}$Institute of Particle and Nuclear Physics, Charles University, Prague, Czech Republic.\\
$^{18}$Nuclear Physics Laboratory, University of Washington, Seattle, WA, USA.\\
$^{19}$Institute for Nuclear Studies, Warsaw, Poland.\\
$^{20}$Institute for Experimental Physics, University of Warsaw, Warsaw, Poland.\\
$^{21}$Rudjer Boskovic Institute, Zagreb, Croatia.
}}

\begin{document}
\maketitle

\begin{abstract}
  The energy dependence of hadron production in central Pb+Pb
  collisions is presented and discussed. In particular, midrapidity
  $m_T$-spectra for $\pi^-$, $K^-$, $K^+$, $p$, $\bar{p}$, $d$,
  $\phi$, $\Lambda$ and $\bar{\Lambda}$ at 40, 80 and 158~\GeV{} are
  shown. In addition $\Xi$ and $\Omega$ spectra are available at
  158~\GeV. The spectra allow to determine the thermal freeze-out
  temperature $T$ and the transverse flow velocity $\beta_T$ at the three
  energies. We do not observe a significant energy dependence of these
  parameters; furthermore there is no indication of early thermal
  freeze-out of $\Xi$ and $\Omega$ at 158~\GeV. Rapidity spectra for
  $\pi^-$, $K^-$, $K^+$ and $\phi$ at 40, 80 and 158~\GeV{} are shown,
  as well as first results on $\Omega$ rapidity distributions at
  158~\GeV. The chemical freeze-out parameters $T$ and $\mu_B$ at the
  three energies are determined from the total yields.  The parameters
  are close to the expected phase boundary in the SPS energy range and
  above. Using the total yields of kaons and lambdas, the energy
  dependence of the strangeness to pion ratio is discussed.  A maximum
  in this ratio is found at 40~\GeV. This maximum could indicate the
  formation of deconfined matter at energies above 40~\GeV. A search
  for open charm in a large sample of 158~\GeV{} events is
  presented. No signal is observed. This result is compared to
  several model predictions.
\end{abstract}

\section{Introduction}
In the last few years NA49 has taken data at 40, 80 and 158~\GeV{}
beam energy to study the energy dependence of particle production in
nucleus-nucleus collisions and to search for indications of the onset
of deconfinement in these collisions. In this paper recent results on
particle spectra and yields are reported.  Results by NA49 on
other observables such as HBT, fluctuations and system size
dependences from p+p, p+A, and A+A (peripheral Pb+Pb, central Si+Si
and C+C) are presented elsewhere in these proceedings
\cite{blume,kreps,hoehne,fischer}.

% Details of $\Lambda$ and
%$\bar{\Lambda}$ production in central collisions at 40, 80 and 158
%\GeV{} were also presented at this conference~\cite{mischke}.
 
In addition to the energy scan programme, a large dataset of central
Pb+Pb data has been collected at 158~\GeV, allowing to study rare
particles such as $\Omega$, $D$, and $\phi \rightarrow
\mathrm{e^+e^-}$. Results on $\Omega$ and open charm production are
presented here.

\section{Experiment and centrality selection}
The NA49 experimental setup~\cite{Afanasev:1999iu} consists of four
TPCs, two of which are in a magnetic field. The Main TPCs (MTPC) are
outside the magnetic field and are used to identify particles by
measuring the energy loss \dEdx.  The particle identification
capabilities of the MTPCs are augmented by two time-of-flight (TOF)
detector arrays which have acceptance around midrapidity for kaons and
close to midrapidity for protons and deuterons. Centrality selection
of the events is done using a calorimeter which detects the projectile
spectators. At 40 and 80~\GeV{} the 7.2\% most central events were
selected. For the 158~\GeV{} data the on-line centrality cut is at
10\%, but for the analysis of pions, kaons and $\phi$ an off-line cut
was made at 5\%. The high statistics data sample used for the
$\Omega$ analysis at 158~\GeV{} was taken with a centrality trigger at
20\% of the inelastic cross section.

\section{Transverse mass spectra and thermal freeze-out}
\begin{figure}
    \epsfig{file=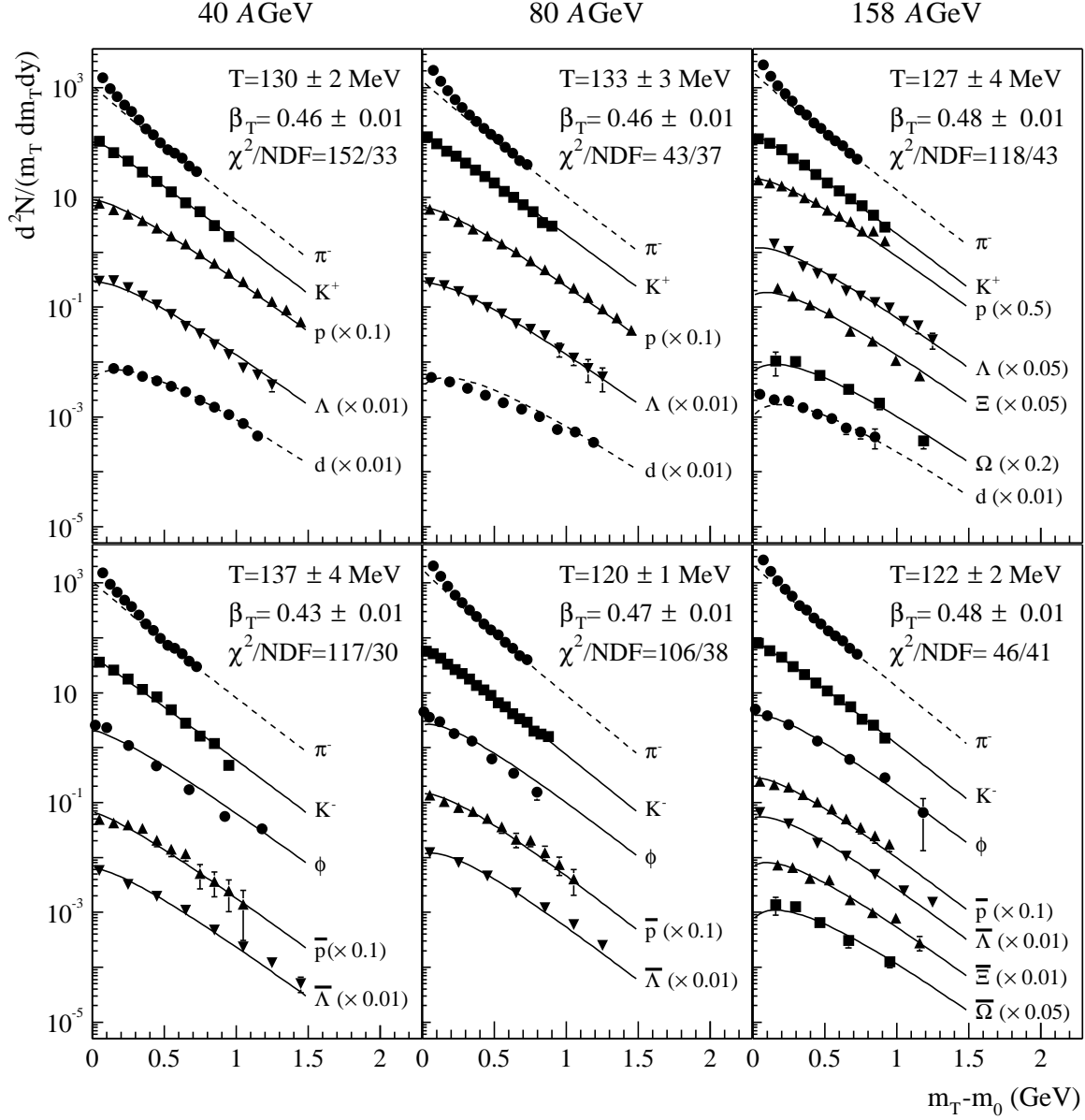,width=\textwidth}
   \caption{Transverse mass spectra in central 40 (left), 80 (middle) and
     158~\GeV{} (right) Pb+Pb
     collisions.
     The lines show the result of the transverse flow fit
     as described in the text.}
   \label{fig:mtspec_all}
\end{figure}
% \begin{figure}
%   \begin{minipage}{0.49\textwidth}
%     \epsfig{file=../blast_wave_40_bw.eps, bb=0 0 510 415, width=\textwidth}
%   \caption{$m_T$-spectra in central 40~\GeV{} Pb+Pb
%     collisions. The lines show the result of the transverse flow fit.}
%   \label{fig:mtspec_40}

%   \end{minipage}%
% \hspace{0.02\textwidth}%
%   \begin{minipage}{0.49\textwidth}
%     \epsfig{file=../blast_wave_80_bw.eps, bb=0 0 510 415, width=\textwidth}
%   \caption{$m_T$-spectra in central 80~\GeV\
%     Pb+Pb collisions. The lines show the result of the transverse flow fit.}
%   \label{fig:mtspec_80}
%   \end{minipage}
% \end{figure}

% \begin{figure}
% \centering
% \epsfig{file=../blast_wave_158_bw.eps, bb=0 0 510 415, width=0.65\textwidth}
% \caption{\label{fig:mtspec_160}$m_t$-spectra measured in central 158~\GeV{} collisions. The lines show the result of the transverse flow fit.}
% \end{figure}

In \fig{fig:mtspec_all}
%Figs.~\ref{fig:mtspec_40}, \ref{fig:mtspec_80} and
%\ref{fig:mtspec_160} 
all available results on $m_T$-spectra as measured in central Pb+Pb
collisions by NA49 are collected.\footnote{The $\bar{p}$, $\Omega^-$ and
  $\bar{\Omega}^+$  $m_T$-spectra at 158 \GeV{} appearing in the printed
  version of these proceedings (Nucl.\ Phys.\ {\bf A715} 161c) are
  unfortunately incorrect and therefore differ slightly from those
  presented here.}
 The spectra are measured at or close
to midrapidity except for the $\phi$, $\Xi$ and $\Omega$ which are
integrated over approximately 1 unit of rapidity, to increase
statistics.

The $\Xi$ and $\phi$ at 158~\GeV{} have already been published
\cite{Afanasiev:2002he,Afanasev:uu} while the kaon and pion spectra are
submitted for publication~\cite{Afanasiev:2002mx}. All other presented
results are still preliminary.

%The $\Lambda$ results were
%presented on this conference~\cite{mischke} and are close to final.
%All other results are newly obtained preliminary results.

In a hydro-dynamical picture, the $m_T$-spectra of
particles are sensitive to transverse flow. To characterise the 
flow, the spectra were fitted with~\cite{Schnedermann:gc}
\begin{equation}
\label{eq:blast_wave}
\frac{dN}{m_Tdm_Tdy} \propto m_T K_1\left(\frac{m_T \cosh\rho}{T}\right)
I_0\left(\frac{p_T \sinh\rho}{T}\right).
\end{equation}
A combined fit of several spectra with this function allows to
determine the thermal freeze-out temperature $T$ and the mean
transverse flow velocity $\beta_T$ ($\rho={\rm atanh}\;\beta_T$). The
$\pi^-$ and deuteron spectra were excluded from the fit (dashed lines
in \fig{fig:mtspec_all}) because the pions are expected to have a
significant contribution from resonance decays and the deuterons may
be formed by coalescence. The particles (baryons and $K^+$) and
anti-particles (anti-baryons and $K^-$) were fitted separately. The
$\phi$ was included in the anti-particle fit, because it is not
expected to be sensitive to the baryon density. The fit describes the
data reasonably well, even including the pions at higher $m_T$ and the
deuterons. The relatively large $\chi^2$ per degree of freedom can
partly be explained by the fact that only statistical errors were
taken into account in the fit (systematic errors are not yet available
for all spectra). Since the slopes of the $m_T$-spectra of particles
and anti-particles are similar, the differences between the freeze-out
parameters for both groups of particles are small.

The differences between the three different energies are not larger
than the difference between the particles and anti-particles at each
energy, implying that the energy dependence of the thermal freeze-out
conditions is small at these energies.

The $\Xi$ and $\Omega$ spectra, which are only
available at 158~\GeV, do not show large deviations from the fit.
Thus, based on these data, there is no indication of an early freeze-out
as has been suggested by several
authors~\cite{vanHecke:1998yu,Dumitru:1999sf}. The slope parameters of
the $\Omega$ spectra are compatible with earlier measurements by WA97
\cite{Antinori:je}.

\section{Rapidity spectra and baryon density}
\begin{figure}
\centering
\epsfig{file=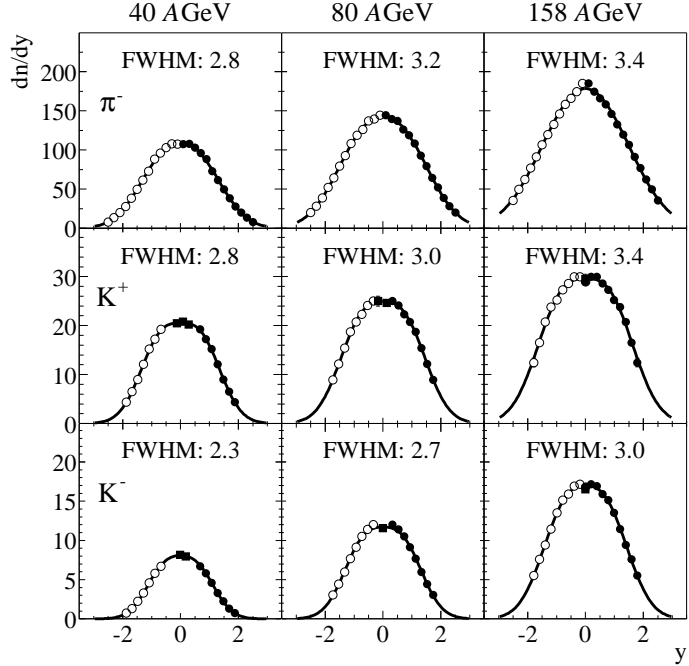, width=0.6\textwidth}
\caption{\label{fig:pika_spec}Rapidity spectra for kaons and
  pions at three different energies. The open points are reflected around midrapidity.
  For kaons, the circles indicate the result of the \dEdx{} analysis,
  while the squares are the result of the combined \dEdx-TOF
  analysis. The distributions are fitted with a double Gaussian and the
  resulting full width at half maximum (FWHM) is indicated in the figure.}
\end{figure}
The $m_T$-spectra of kaons as shown in \fig{fig:mtspec_all} were
obtained using combined \dEdx{} and TOF information at midrapidity.
The \dEdx{} measurement in the TPC alone allows to identify kaons at
forward rapidities. The resulting spectra are shown in \fig{fig:pika_spec},
together with $\pi^-$ distributions which were obtained using unidentified
negatively charged particles, corrected for contributions of $K^-, \bar{p}$ and
non-vertex tracks.

The $K^-$ distributions are narrower than the $K^+$ distributions at
all energies. This could indicate that the kaons are sensitive to the local
baryon density, which is reflected in the 
difference between the $\Lambda$ and $\bar{\Lambda}$ spectra
 as well (for details on the $\Lambda$ analysis, see~\cite{mischke} in
 these proceedings).
 
\begin{figure}
  \epsfig{file=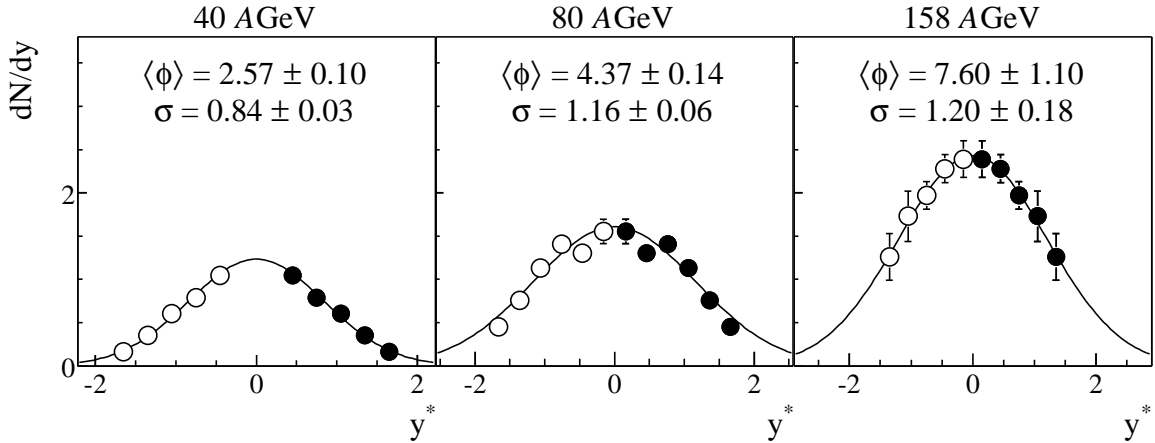,width=\textwidth}
\caption{\label{fig:phi}Rapidity spectra of $\phi$ production at 40,
  80 and 158~\GeV{} respectively. The open points are reflected around
  mid-rapidity. The full lines show a Gaussian fit, used to obtain the
  mean multiplicities in full phase space $\langle\phi\rangle$. The
  fitted widths $\sigma$ are given as well.}
\end{figure}
The $\phi$ spectra are obtained from invariant mass distributions of
$K^+K^-$ pairs. The resulting rapidity spectra are shown in
\fig{fig:phi}, where the new results at 40 and 80~\GeV{} are compared
to the published result at 158~\GeV{}~\cite{Afanasev:uu}. Clearly, the
width of the distribution, as well as the total yield, increases with
beam energy.

\begin{figure}
\begin{minipage}{0.6\textwidth}
\epsfig{file=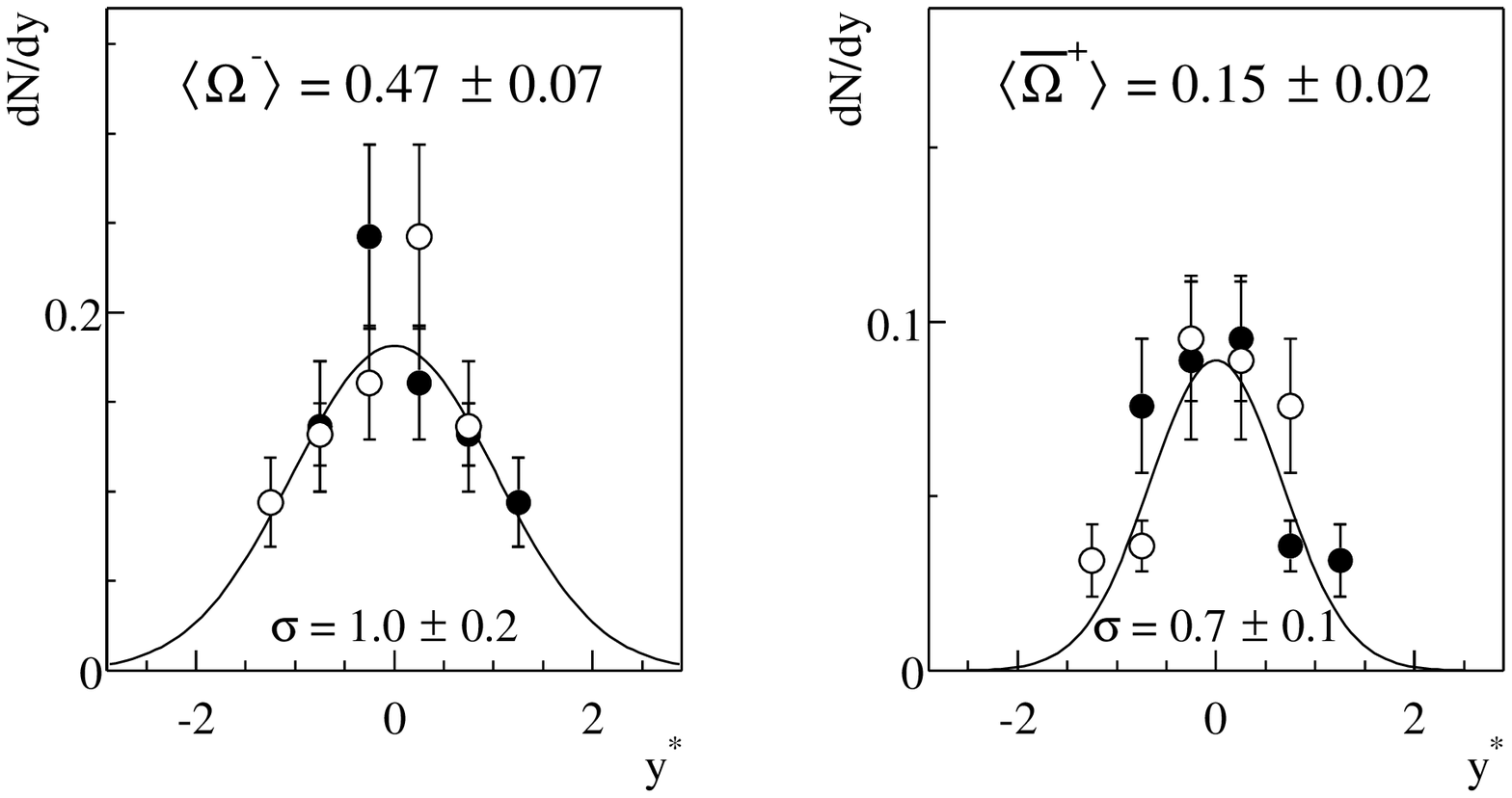,width=\textwidth,bb=10 20 540 300}
\caption{\label{fig:omega_rap}Rapidity spectra of \omeg{} and \aomeg{} at
  158~\GeV. The total yields $\langle\Omega\rangle$ were calculated by
  integrating the fitted Gaussians (full line). The widths $\sigma$ of
  the fitted curves are also given.}
%\vspace{0.9cm}
\end{minipage}
\hspace{0.02\textwidth}
\begin{minipage}{0.38\textwidth}
  \epsfig{file=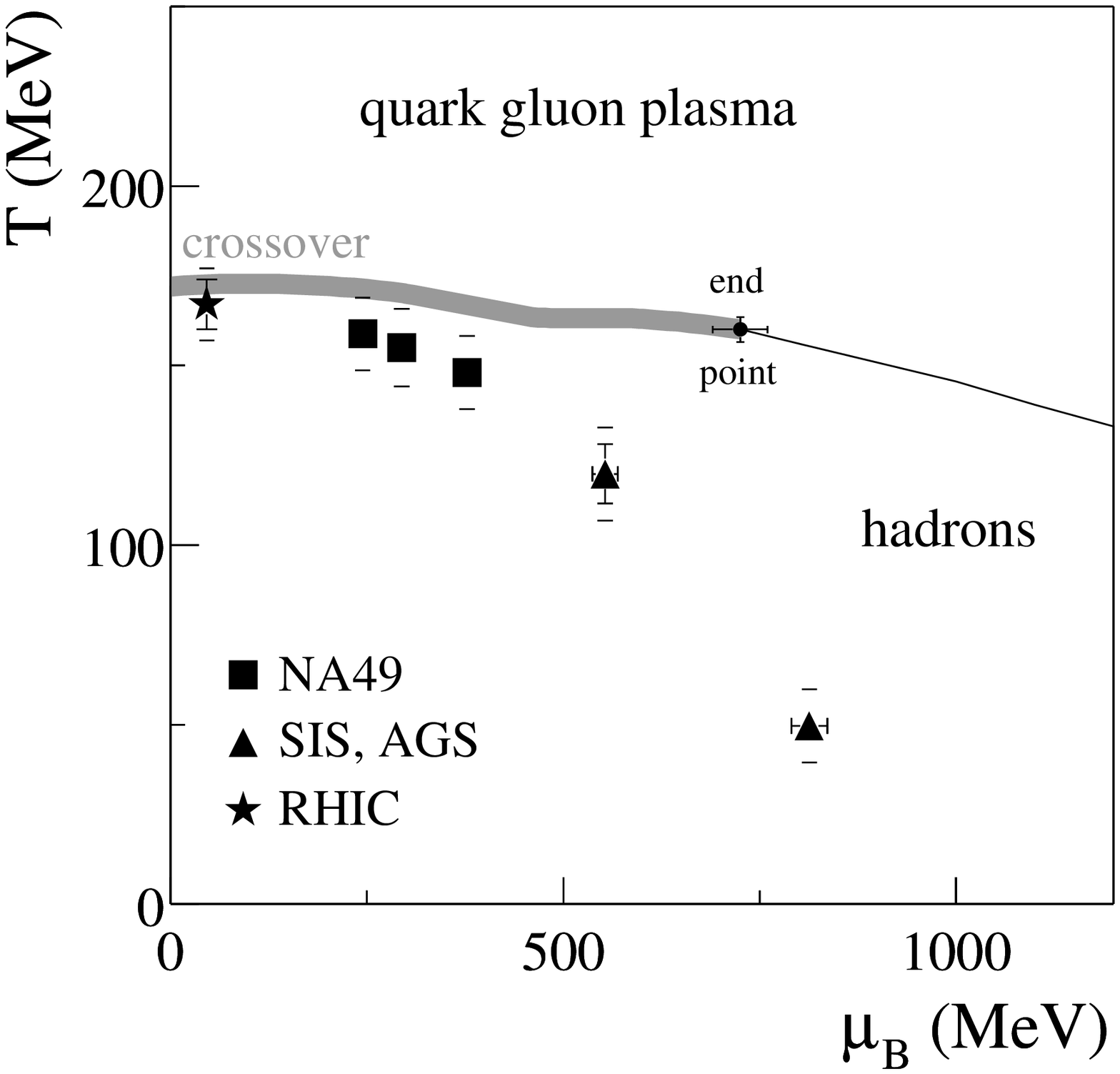, width=\textwidth}
  \caption{\label{fig:chem_pars}Chemical freeze-out points as
  determined from a hadron gas fit at different energies~\cite{becattini}. The
  crossover curve is taken from~\cite{Fodor:2001pe}.}
\end{minipage}

\end{figure}
The large statistics data sample of the 20\% most central Pb+Pb
collisions at 158~\GeV{} beam energy, which was taken in the year 2000, has
been used for the $\Omega$ analysis. The resulting $m_T$-spectra are
shown in \fig{fig:mtspec_all} and the corresponding rapidity spectra
in \fig{fig:omega_rap}. Integrating over rapidity gives mean
multiplicities per event of $\langle \Omega^- \rangle=0.47\pm0.07$ and
$\langle \bar{\Omega}^+ \rangle=0.15\pm0.02$. The observation that the
ratio $\aomeg/\omeg$ is below unity cannot be explained in a pure
string fragmentation model but fits to expectations in an equilibrium
hadron gas model~\cite{Bass:2002fx}.  Also note that the \omeg{} has a
slightly wider rapidity distribution than the \aomeg.  This
indicates that even the $\Omega$ is sensitive to the effect of baryon
density.

\section{Total yields and chemical freeze-out}
The measured rapidity spectra have been integrated to obtain the mean
multiplicities in full phase space. To determine the chemical
freeze-out parameters at the three different energies, the total
yields were fitted with a hadron gas model with partial strangeness
saturation~\cite{becattini}. The obtained temperatures for chemical
freeze-out (145--160~MeV) are larger than the thermal freeze-out
temperatures (115--140~MeV). The chemical freeze-out temperature $T$
and baryon chemical potential $\mu_B$ are compared to results at AGS
and RHIC in \fig{fig:chem_pars}. The freeze-out points follow a smooth
curve of increasing temperature and decreasing baryon density with
energy. The observed freeze-out parameters are close to the calculated
phase boundary~\cite{Fodor:2001pe} at SPS energies and above.

\section{Energy dependence of strangeness production}
\begin{figure}
\centering
\epsfig{file=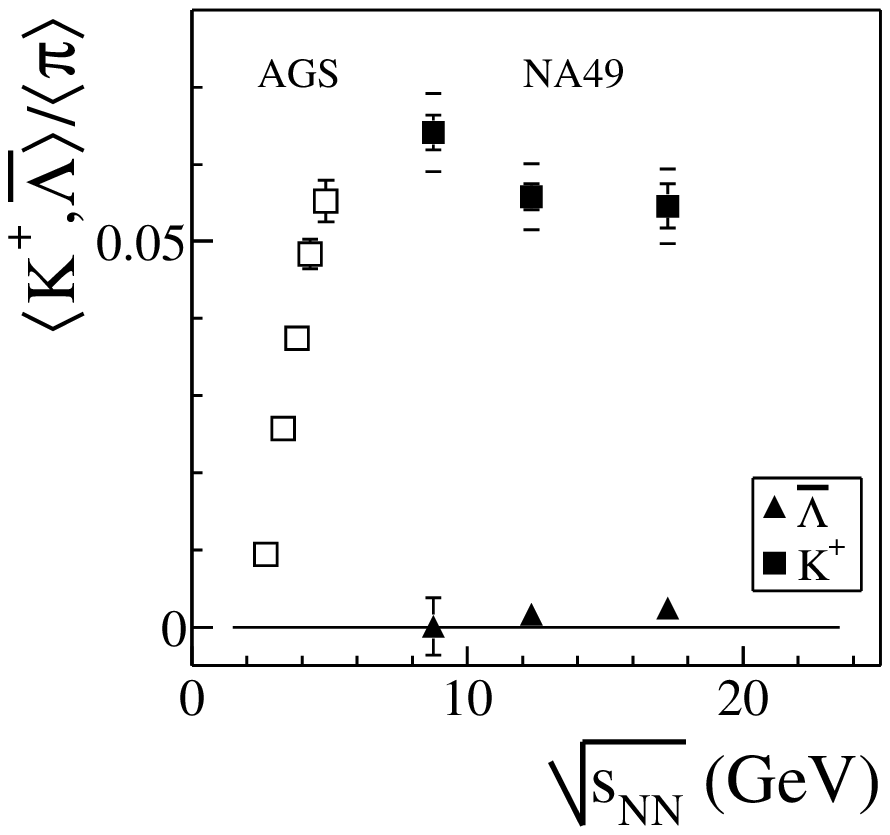,width=0.45\textwidth}
\epsfig{file=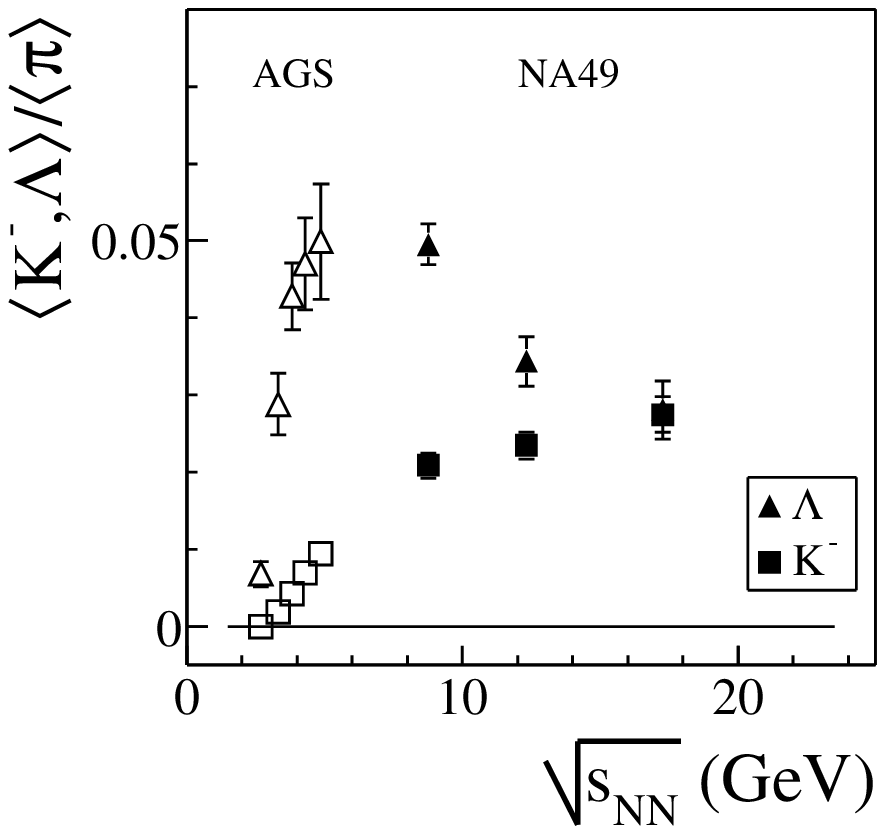,width=0.45\textwidth}
\caption{\label{fig:kapi}
  Total yields of $K^+$ and $\bar{\Lambda}$ (left) and $K^-$ and
  $\Lambda$ (right), normalised to the total pion multiplicity,
  as a function of $\sqrt{s}$ for central Pb+Pb (NA49) and Au+Au (AGS)
  collisions.}
\end{figure}

\begin{figure}
\begin{minipage}{0.48\textwidth}
\epsfig{file=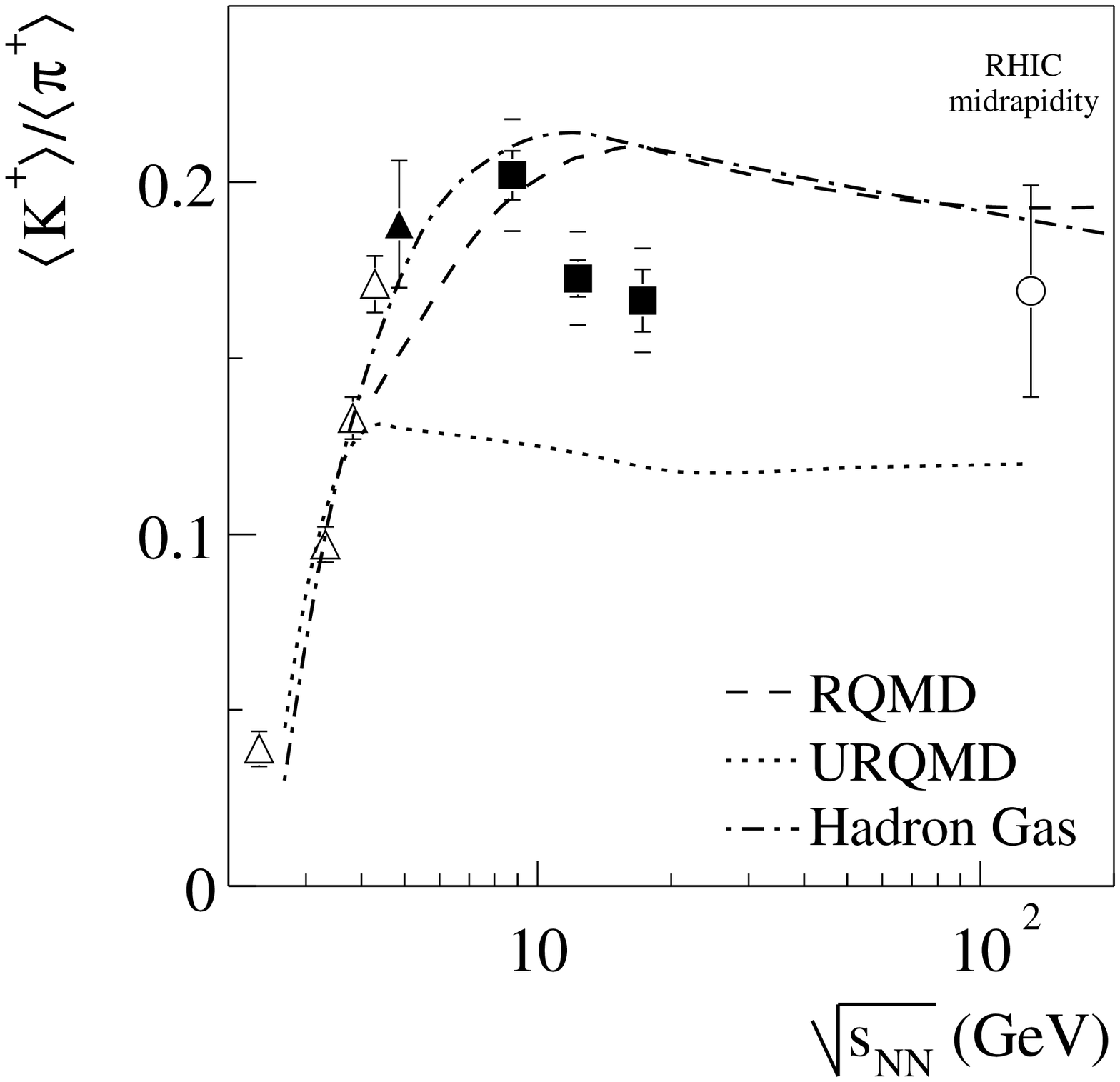, width=\textwidth}
\caption{\label{fig:kapi_plus_model}Energy dependence of $\langle K^+
  \rangle/\langle \pi \rangle$ compared to different model predictions
  which do not invoke QGP formation (see text).}
\end{minipage}%
\hspace{0.04\textwidth}
\begin{minipage}{0.48\textwidth}
\epsfig{file=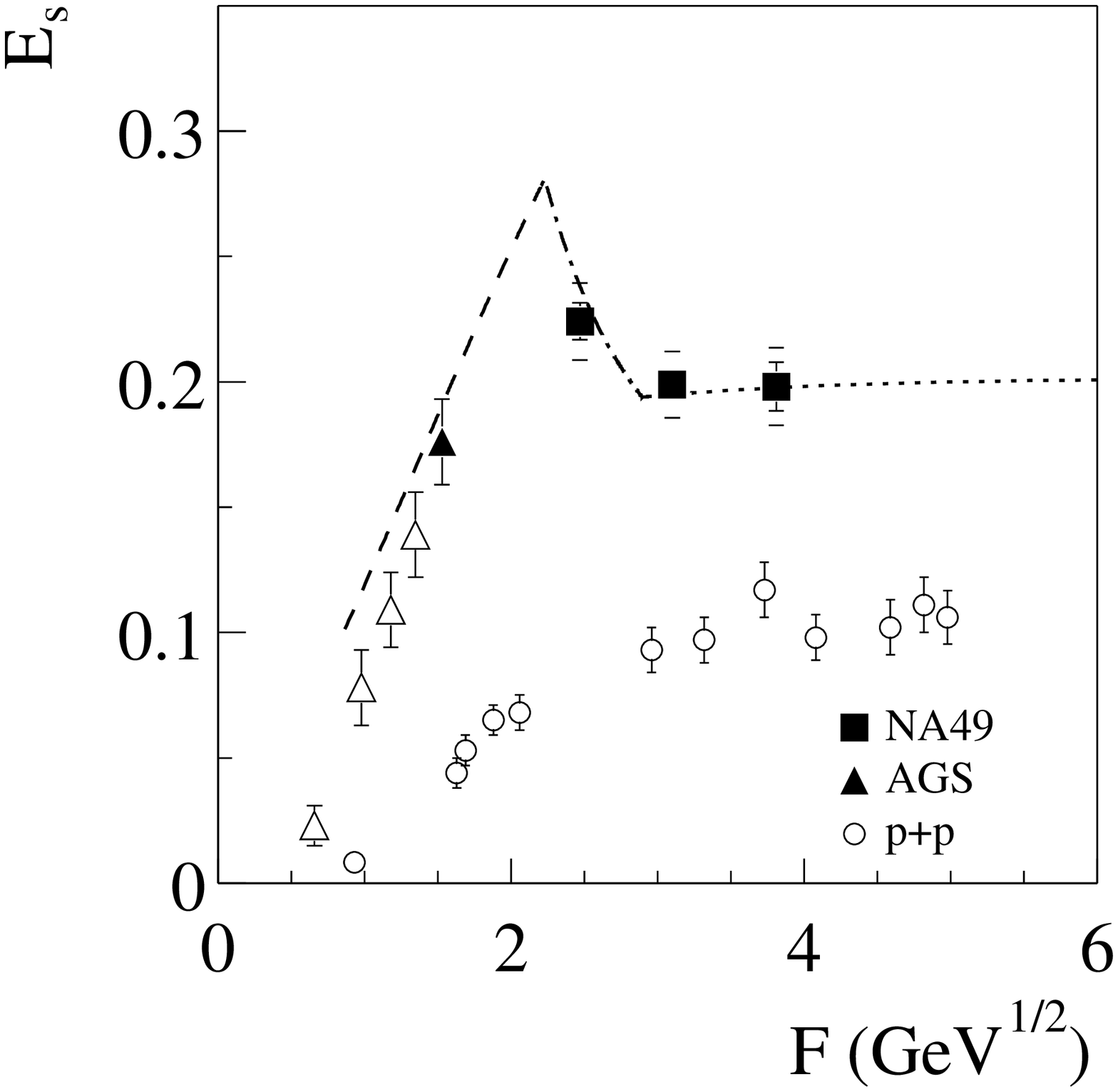, width=\textwidth}
\caption{\label{fig:es}Comparison of the strangeness to entropy ratio
  $E_S$ to a model calculation~\cite{Gazdzicki:1998vd} which does
  assume that a QGP is formed above approximately 30~\GeV.}
\end{minipage}
\end{figure}
The mean multiplicities of the most abundant carriers of strangeness,
kaons and lambdas, are plotted as a function of the centre of mass
energy in \fig{fig:kapi}. Only the charged kaons are shown because the
neutral kaons are not measured at all energies. All yields are divided
by the total pion yield $\langle\pi\rangle=1.5
(\langle\pi^+\rangle+\langle\pi^-\rangle)$. It is seen that the
$\langle K^+\rangle/\langle \pi \rangle$ ratio peaks at 40~\GeV,
whereas $\langle K^-\rangle/\langle \pi \rangle$ increases
monotonically. On the other hand, the $\Lambda$ production also peaks
around 40~\GeV, whereas the $\bar{\Lambda}$ yield increases
monotonically with energy. Apparently, both the total $s$ production
and the $\bar{s}$ production have a maximum close to 40~\GeV. The
maximum in the $\bar{s}$ production is directly seen $K^+/\pi$ ratio
because the contribution of $\bar{\Lambda}$ is small. Although the
$K^-$ represent a large fraction of the $s$ production, the maximum in
the $s$ production is reflected in the $\Lambda/\pi$ ratio.

In \fig{fig:kapi_plus_model} the energy dependence of the $\langle
K^+\rangle/\langle\pi^+\rangle$ ratio is compared to several models
which do not assume the formation of a QGP in the collision. The UrQMD
model~\cite{urqmd} fails to describe the data because the pion yields
are over-estimated by about 30\%. The RQMD model~\cite{rqmd} gives a
much better descriptions, but over-estimates the ratios at 80 and
158~\GeV. The same is true for the hadron gas
model~\cite{Braun-Munzinger:2001as}. In this model, the $\langle
K^+\rangle/\langle\pi^+\rangle$ ratio peaks around 40~\GeV{} due to
the interplay of the decreasing baryon density and the increasing
temperature with energy.

In \fig{fig:es} the energy dependence ($F \equiv
(\sqrt{s}-2m_N)^{3/4}/\sqrt{s}^{1/4}$) of the strangeness to entropy
ratio $E_S=(\langle\Lambda\rangle + \langle K \rangle +
\langle\bar{K}\rangle)/\langle \pi \rangle$ is compared to a
model~\cite{Gazdzicki:1998vd} which assumes that above a certain
energy a QGP is formed in the collision. The peak in the model curve
is caused by this transition. In~\cite{Afanasiev:2002mx} it is shown
that the model also describes the energy dependence of pion production.

NA49 intends to complete the energy scan by taking data at 20 and
30~\GeV{} in 2002.

\section{Open charm upper limit}
\begin{figure}[b!]
\epsfig{file=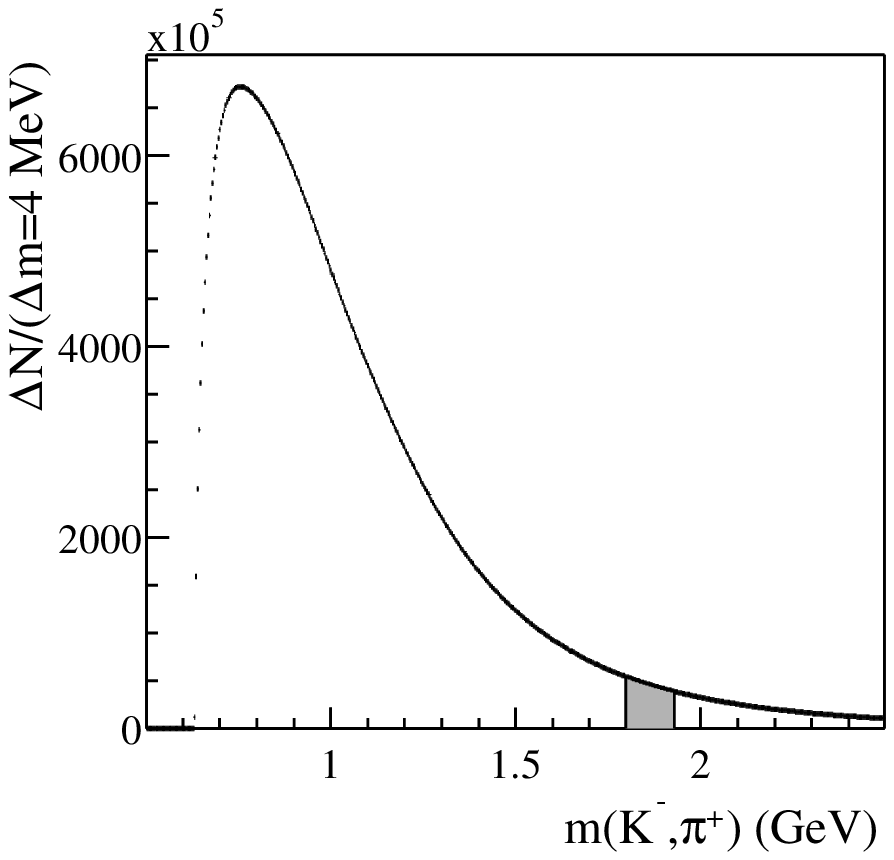, bb=0 10 283 240, width=0.48\textwidth}
\hfill
\epsfig{file=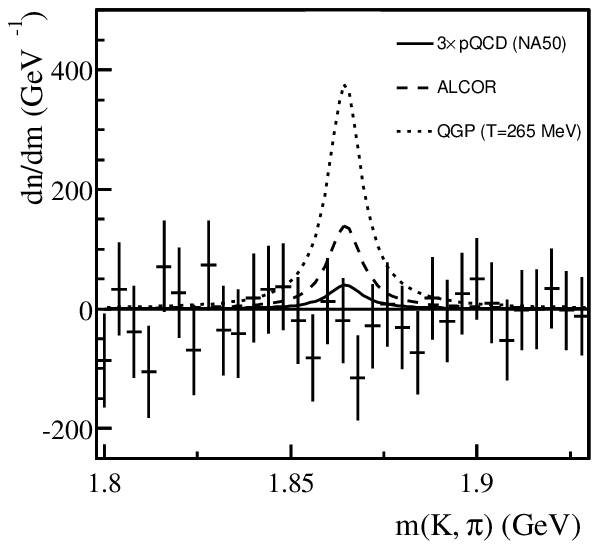, bb=378 7 567 160, width=0.48\textwidth}
\caption{\label{fig:d0_imass}Uncorrected invariant mass distribution of
  all $K^-,\pi^+$ pairs (left) and invariant mass distribution for the
  sum of all $D^0$ and $\bar{D}^0$ candidates after background
  subtraction (right), corrected for acceptance and efficiency. The
  curves indicate predictions of various models (see text).}
\end{figure}

All available central Pb+Pb data at 158~\GeV{}
(900k events at 10\% and 3M events
at 20\% centrality) were used in an invariant mass
analysis to measure open charm. The invariant mass was calculated for pairs of pion and kaon candidates, to detect the decays
\[
\label{eq:d0_decay}
D^0 \rightarrow K^- + \pi^+ \mathrm{~~and~~} 
\bar{D}^0 \rightarrow K^+ + \pi^-,
\]
which have a branching fraction of 3.83\%. For tracks which reach the
MTPC, the kaon candidates were selected by \dEdx{} measured in the
MTPC. For tracks which do not reach the MTPC, all candidates were
used. The invariant mass histogram for $D^0$ before background
subtraction is shown in the left panel of \fig{fig:d0_imass}. The
result after adding the $\bar{D}^0$, subtracting the background and
correcting for acceptance and efficiency losses is shown in the right
panel of \fig{fig:d0_imass}. No signal of the
decays is visible. Detector simulations have shown that the shape of
the expected signal can be parametrised as a Breit-Wigner with a width
$\Gamma=11$~MeV. Using this parametrisation, predictions by different
models are shown in the figure. It is seen that thermal production of
charm quarks in a Quark Gluon Plasma at $T=265$~MeV, as described in
\cite{Gazdzicki:1999mc}, is excluded (dotted curve in
\fig{fig:d0_imass}). The expectation from the quark coalescence model
ALCOR~\cite{Levai:2000ne} is at the limit of the sensitivity of the
present analysis (dashed curve). Lower values of the
charm yield, such as the expectation from perturbative QCD, or a
factor 3 enhancement~\cite{Abreu:2000nj} (full curve) cannot
be excluded. In order to quantify the result, the data have been
fitted with the parametrised peak shape, leaving the normalisation as
the only free parameter. The result is $\langle D^0 +
\bar{D}^0\rangle=-1.07 \pm 0.84$.

\section{Summary and outlook}
Spectra of $\pi, K, p, d, \phi$, and $\Lambda$ in central Pb+Pb
collisions at 40, 80 and 158~\GeV{} are presented, together with new results on
$\Omega$ rapidity and $m_T$ spectra at 158~\GeV.

The $m_T$-spectra are compatible with radial flow, with similar
temperature $T\approx130$~MeV and transverse flow velocity
$\beta_T\approx0.45$  at the different
energies. The $\Xi$ and $\Omega$ spectra, which are only measured at 158~\GeV{}, do
not show indications of early freeze-out.

The chemical freeze-out parameters as determined from the measured
total yields approach the phase boundary in the SPS energy range.

The energy dependence of strangeness production shows a maximum in the
relative strangeness yield at 40~\GeV. This maximum is observed in the
$K^+/\pi$ ratio for the $\bar{s}$ and in the $\Lambda/\pi$
ratio for the $s$ quarks. The maximum can be explained by the onset
of deconfinement in the energy range between AGS and SPS. Upcoming
runs at 20 and 30~\GeV{} will provide a more accurate localisation of
the maximum.

To investigate open charm production in heavy ion collisions, an
invariant mass analysis has been performed using a large sample of
158~\GeV{} Pb+Pb events. No signal has been observed. The result
excludes equilibrium charm production in a QGP of $T=265$~MeV.

\paragraph{Acknowledgements}
We would like to thank F. Becattini for providing the results
of his hadron gas fits.

This work was supported by the Director, Office of Energy Research, 
Division of Nuclear Physics of the Office of High Energy and Nuclear Physics 
of the US Department of Energy (DE-ACO3-76SFOOO98 and DE-FG02-91ER40609), 
the US National Science Foundation, 
the Bundesministerium fur Bildung und Forschung, Germany, 
the Alexander von Humboldt Foundation, 
the UK Engineering and Physical Sciences Research Council, 
the Polish State Committee for Scientific Research (5 P03B 130 23 and 2 P03B 02418), 
the Hungarian Scientific Research Foundation (T14920 and T32293),
Hungarian National Science Foundation, OTKA, (F034707),
the EC Marie Curie Foundation,
and the Polish-German Foundation.

\end{document}